\documentclass{ws-ijmpa}

\newcommand{\eq}{\begin{equation}}
\newcommand{\eqx}{\end{equation}}
\newcommand{\eqn}{\begin{eqnarray}}
\newcommand{\eqnx}{\end{eqnarray}}
\newcommand{\nc}{$ N_{\rm cut}\  $}

\chardef\TL=`\~ 
\chardef\UL=`\_ 

\begin{document}

\markboth{Jacek Wosiek} {Supersymmetric Yang-Mills quantum
mechanics ... }

%
\catchline{}{}{}{}{}
%

\title{SUPERSYMMETRIC YANG-MIILS QUANTUM MECHANICS IN VARIOUS
DIMENSIONS\footnote{Presented at the Eighth Workshop on Non-Perturbative QCD, Paris, June 2004.}
}

\author{\footnotesize JACEK WOSIEK 
}

\address{M. Smoluchowski Institute of Physics, Jagellonian University, \\
Reymonta 4, 30-056 Krakow, Poland 
 }

\maketitle


\begin{abstract}
Recent analytical and numerical solutions of above systems are reviewed.
Discussed results include: a) exact construction of the supersymmetric vacua
in two space-time dimensions, and b) precise numerical calculations of the coexisting, continuous and discrete,
spectra in the four-dimensional system, together with the identification of dynamical supermultiplets
and SUSY vacua.
New construction of the gluinoless SO(9) singlet state,  which is vastly different from the
empty state, in the ten-dimensional model is also briefly summarized.
\end{abstract}
%
\noindent TPJU-16/2004\newline
MPP-2004-125

\section{Introduction}    
Supersymmetric Yang-Mils quantum mechanics (SYMQM) results from the dimensional reduction
of the corresponding field theory to a single point in the D-1 dimensional space.
Equivalently, it can be thought of as the effective quantum mechanics of zero momentum modes
of the latter~\cite{BSS}.
These systems were first considered in 80's~\cite{CH} as simple models with supersymmetry~\cite{WI}.
Independently, zero-volume field theories (especially pure Yang-Mills) were being used as the starting
point of the small volume expansion - an important theoretical development complementary to early lattice
calculations~\cite{L,LM,VABA}. In late 90's the models attracted a new wave of interest after
the remarkable hypothesis of the equivalence, between the  $D=10, SU(\infty)$ SYMQM
and M-theory of D0 branes, has been formulated ~\cite{BFSS,BS,PR,BERS,WAT,HS}.

 We review some recent progress in studying the SU(2) models in two, four and ten space-time dimensions.

\section{Exact supersymmetric vacuum in two space-time dimensions}
The system, reduced from $D=2$ to one (time) dimension
\cite{CH}, is described by the three real bosonic variables
$x_a(t)$ and three complex, fermionic degrees of freedom
$\psi_a(t)$, both in the adjoint representation of SU(2) ,
$a=1,2,3$.

The Hamiltonian reads
 \eq H={1\over 2} p_a p_a +i g
\epsilon_{abc} \psi_a^{\dagger} x_b \psi_c ={1\over 2} p_a p_a + g x_a G_a ,
\label{HD2}
\eqx
 where
the quantum operators $\psi,\psi^{\dagger},x,p$
can be written
in terms of the creation and annihilation operators
\eq
   x_a={1\over\sqrt{2}}(a_a+a_a^{\dagger}),\;\;  p_a={1\over i \sqrt{2}}(a_a-a_a^{\dagger}), \;\; \label{XPD2}
   \psi_a=f_a,\;\;  \psi_a^{\dagger}=f_a^{\dagger},
\label{eq:f} \eqx
and gauge generators $G_a$ read
\eq G_a=\epsilon_{abc}(x_b p_c - i \psi_b^{\dagger}
\psi_c). \label{GD2} \eqx
The physical Hilbert space
consists only of gauge invariant states. This constraint is accommodated by constructing all possible
  invariant combinations of
creation operators (creators), and using
them to generate a complete gauge invariant basis of states. There
are four lower order creators: \eq (aa)  \equiv  a_a^{\dagger}
a_a^{\dagger},\;\; (af)  \equiv  a_a^{\dagger} f_a^{\dagger},\;\;
(aff) \equiv  \epsilon_{abc}a_a^{\dagger}
f_b^{\dagger}f_c^{\dagger},\;\; (fff) \equiv
\epsilon_{abc}f_a^{\dagger} f_b^{\dagger}f_c^{\dagger}.
\label{creators} \eqx
In the physical basis the Hamiltonian (\ref{HD2}) reduces to that
of a free particle. Consequently the gauge invariant
fermion number $F=f_a^{\dagger} f_a$ is conserved -
the whole basis
splits into the four sectors, each sector beginning with one of the
following "base" states
\eq |0_F\rangle=|0\rangle,\;\;
|1_F\rangle=(af)|0\rangle,\;\; |2_F\rangle=(aff)|0\rangle,\;\;
|3_F\rangle=(fff)|0\rangle. \label{ground}
\eqx
To generate the whole basis in is then sufficient to act repeatedly on these four vectors
with the bosonic creator $(aa)$ \footnote{Notice that $ (ff)=(af)^2=(aff)^2=(fff)^2=0 $.}.
 The
basis with a cutoff \nc is then obtained by applying $(aa)$ up to
\nc times. Obviously our cutoff is gauge invariant, since it
is defined in terms of the gauge invariant creators.

In Refs.~\cite{JW1,CW1} such a basis was explicitly constructed in the "Mathematica representation" of
quantum mechanics following by the numerical diagonalization
of the Hamiltonian matrix for finite but large cutoffs. Resulting spectrum is indeed that of a free
particle regularized in a harmonic oscillator basis. Eigenstates obtained in this way
 are in close correspondence with the gauge invariant plane waves of Ref.~\cite{CH}, $ r^2=x_a x_a $,
\eqn
|0_F,k>={\sin{(kr)}\over kr} |+>,&\;\;&
|1_F,k>=\left( {\sin{(kr)}\over (kr)^2} - {\cos{(kr)}\over kr} \right) {(x_a f^{\dag}_a)\over r}|+>,\\
|3_F,k>={\sin{(kr)}\over kr} |->,&\;\;&
|2_F,k>=\left( {\sin{(kr)}\over (kr)^2} - {\cos{(kr)}\over kr} \right) {(x_a f_a)\over r}|->.
\eqnx
Where $|\pm>$ were referred to as the "empty" and "filled" states.
In the limit $k\rightarrow 0$ the second and fourth state
vanish, which confirms that the empty (and filled) states are
the (unpaired) supersymmetric vacua of the model. It is important to keep in mind that, e.g., the $|+>$  state
{\em is not} empty in terms of the bosonic quanta.  Obviously the $|0_F,0_B>$
 it is even not an eigenstate of $H$.
On the other hand, we can easily construct the exact vacuum state(s) using results
of the recursive approach of Ref.~\cite{CW1}. The Hamiltonian matrix in the $F=0$ basis is tridiagonal
\eq
<2n|H|2n-2>=<2n|H|2n-2>=-{1\over 4}\sqrt{2n+4n^2},\;\;
<2n|H|2n>=n+{3\over 4},
\eqx
and consequently the eigenequation
$ H|+>=0|+> $
turns over into the recursive relation for the components of the vacuum state in the $F=0$ basis
\eq
\sqrt{(2n+2)(2n+3)}c_{2n+2}-(4n+3)c_{2n}+\sqrt{2n(2n+1)}c_{2n-2}=0, \label{rrel}
\eqx
which determine the vacuum state
\eq
|+>=\Sigma_{n=0}^{\infty} c_{2n}|0_F,2n>,
\eqx
up to a normalization. Due to the particle-hole symmetry, similar equation holds for the $|->$
state in the $3_F$ sector.

Moreover, given the explicit representation of the supersymmetry generator $Q$ in the $F=0$ basis
\cite{CW1}, one readily verifies that it annihilates the above vacuum state
  \eq
  Q_{mn}c_{n}=0
  \eqx
as it should. Therefore the "empty" and "filled" qualifiers in Ref.~\cite{CH} refer to the fermionic
occupation numbers only. In terms of the bosonic quanta both supersymmetric vacua have the nontrivial
structure given by Eq.(\ref{rrel}).

\section{Dynamical supermultiplets in four space-time dimensions}
Four-dimensional system is of course much more rich and correspondingly more interesting.
No analytical solutions are known in spite of various attempts \cite{CH}. There are now 15
degrees of freedom which upon quantization are represented by nine bosonic $a_b^{i\dag}$ and six
fermionic $f_a^{m\dag}$ creation operators (see Ref.~\cite{JW1} for the details of the construction).
Fermion number is again conserved which allowed separate diagonalization in each of seven $F=0,1,..,6$
fermionic sectors. The gauge and rotationally invariant cutoff is again introduced as the maximal number
of all bosonic quanta $B_{max}$. Direct implementation of the cut basis was practical only for
 $ B_{max}\le 8$
and resulted in several thousands of all basis vectors. It was sufficient to perform the satisfactory
comparison with earlier results known in the $F=0$ sector~\cite{LM} which is identical with the pure Yang-Mills system.
Present results are obtained
for much higher cutoffs, in the range of $30 - 60$, in all fermionic sectors ~\cite{CW2}. This progress
was achieved
by taking full advantage of rotational symmetry together with the generalization of the recursive
construction of matrix elements, developed originally for the D=2 model~\cite{CW1}. Second dedicated method
employs the separation of variables done first by Savvidy~\cite{SAV} and applied to the present system by
van Baal~\cite{VBN}.
It allows to reach the highest cutoffs up to date, but was done only for the two channels: $F=0,2$,
both with the lowest total angular momentum $J=0$ . On the contrary the former approach
works for all F's and J's.
This gives now not only eigenenergies and eigenstates, but allows for the precise
identification of supermultiplets and complete classification of the whole spectrum.

\begin{figure}
\centerline{\psfig{file=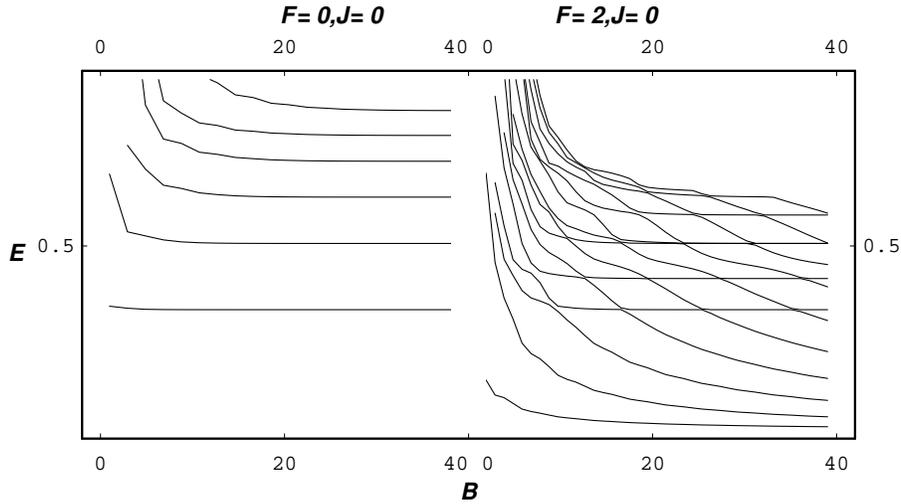,width=12cm}} \vspace*{8pt}
\caption{Cutoff dependence of the spectrum in the zero- and two-gluino sectors.}
\label{vBa}
\end{figure}

Fig.~\ref{vBa} shows the cutoff dependence of the spectrum in the $F=0,J=0$ and $F=2,J=0$ channels,
obtained by separation of variables. One clearly sees the two families of states which differ dramatically
by their convergence with the cutoff. Quickly convergent levels correspond to the discrete spectrum
of localized states, while the slowly falling ones are the non-localized, scattering states
which in the infinite cutoff limit form a continuous spectrum. This result nicely confirms the general
expectations for the behavior of supersymmetric systems with the flat valley potentials \cite{}.
The continuous spectrum exists only in the "fermion rich" sectors supporting the general argument
based on the fermion-boson cancellation of the zero-point energy. In the $F=2$ sector
both the discrete and continuous spectra coexist at the same energy, which is unusual but
not excluded \cite{SIM}. Moreover, the mixing between the discrete states and the continuum vanishes
at infinite cutoff, rendering former stable as required by supersymmetry. Actual energies of the scattering
states are encoded in the rate of fall with the cutoff seen in the Figure. They can be readily recovered
in the scaling limit where the principal quantum number $n$ varies with the cutoff $B$ as
$n\sim\sqrt{B}$ \cite{TW}.

\begin{figure}
\centerline{\psfig{file=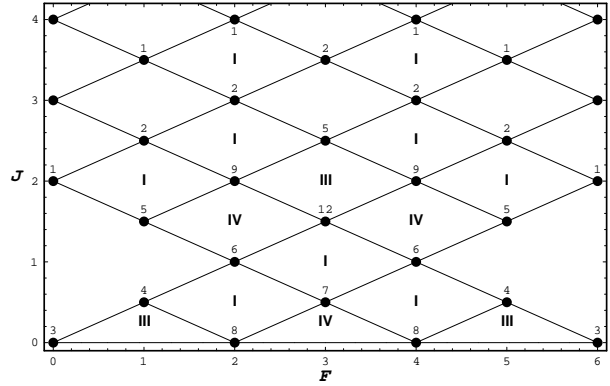,width=8cm}} \vspace*{8pt}
\caption{Sizes of bases vs. numbers of associated supermultiplets for $B_{max}=3$}
\label{MAP}
\end{figure}

Similar results are now available in all channels of angular momentum $J$ and fermion
number $F$ \cite{CW2} \footnote{The cutoffs reached with this approach have been recently pushed
yet further and almost match precision obtained by separating variables \cite{CB}.}.
This allowed rather detailed study of the supersymmetry induced relations among eigenstates
and complete classification of the spectrum into dynamical supermultiplets. The problem is already seen
in Fig.~\ref{vBa}: the two lowest energies in the $F=0,J=0$ channel, $(0_F,0_J)$, clearly agree with
the first and third
discrete levels of $(2_F,0_J)$. Naturally one suspects the supersymmetry to cause this degeneracy.
However the second level of $(2_F,0_J)$ does not have a counterpart in the $(0_F,0_J)$  channel.
Therefore its superpartner must lie in yet another channel.

     To illustrate the systematic study of dynamical supermultiplets it is helpful to consult
a map of the Hilbert space shown in the $F,J$ plane, Fig.~\ref{MAP}. Dots represent specific $(F,J)$
channels, while the adjacent Arabic numbers give sizes of  bases
which can be constructed at given cutoff (here $B_{max}=3$). Notice that, equivalently, one could
redistribute these "occupation numbers" into diamonds attached to each vertex.  In this way we
have generated the "occupation numbers" of diamonds
denoted by roman numbers in the Figure. Each state can belong only to one diamond, therefore we have an array
of local relations
\eq
d_i=\Sigma_{I|i} R_I,
\label{eq:d-vs-R}
\eqx
where $R_I$ denotes a multiplicity of a diamond I, $d_i$ is a number of SO(3) multiplets in
a channel $i$, and summation runs over I's adjacent to $i$. Eqs.(\ref{eq:d-vs-R})
provide global constraints on bases in all channels. This situation holds for any odd cutoff
$B_{max}$. The number of new states created with higher cutoff is such that they always fill
the integer number of diamonds. This regularity is the necessary condition for supersymmetry.
In fact above diamonds are nothing but supermultiplets with identical eigenenergies of all four
members of the supermultiplet. In another words: while dots in the Figure are useful for labeling
basis states, the diamonds are suitable for classifying eigenstates of the Hamiltonian.

\begin{figure}
\centerline{\psfig{file=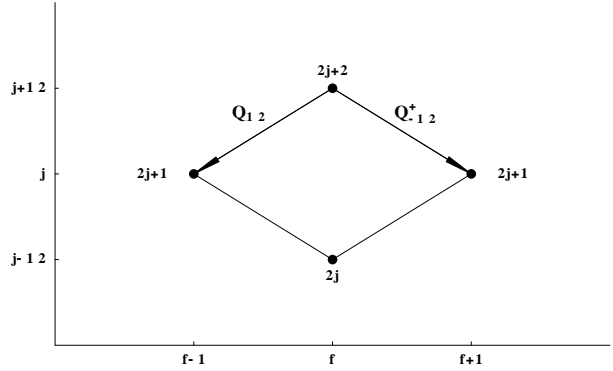,width=8cm}} \vspace*{8pt}
\caption{A single massive supermultiplet described in the text. An example of the action of two
supersymmetry generators is also shown.}
\label{diam}
\end{figure}

    To see that each diamond of Fig.~\ref{MAP} indeed represents a supermultiplet
consider the algebra of Weyl generators $Q_m$ and $Q_m^{\dag}$
\eq
\{Q_m,Q_n^{\dag}\}=2\delta_{mn} H,\;\;\{Q_m,Q_n\}=\{Q_m^{\dag},Q_n^{\dag}\}=0,\;\;   m=\pm {1\over 2}.
\label{susy}
\eqx
Detailed form of these observables is not important here (see Ref.~\cite{CW2} for the details).
It suffices to know that $Q_{m}^{\dag}$ ($Q_{m}$) creates (annihilates) one fermion with the angular
momentum $J_z=m$.
Eq.(\ref{susy}) implies that these generators, when acting on the eigenstates of the Hamiltonian, can only
move them within one diamond, see Fig.~\ref{diam} . Similarly, repeated actions
of up to four Q's transform eigenstates within the same supermultiplet.
Further, for each eigenstate in the spectrum there exists a unique pair of generators which separately
annihilate this state. That pair determines to which of the four neighboring supermultiplets
the state belongs.

It was reassuring to observe how these relations are satisfied by our eigenstates.
To this end we have constructed the {\em supersymmetry fractions}
\eq
 q(j',i'|j,i) \equiv {1 \over 4 E_{j,i}}
   \left|{\langle j';i'\Vert Q^\dagger\Vert j ;i \rangle}\right|^2,
\eqx
which analyze the supersymmetric image of an eigenstate $|F,j;i >$ in another channel $(F+1,j')$.
They do not vanish only if the initial and final states belong to the same supermultiplet.
Since the supersymmetry is restored
only at the infinite cutoff, susy fractions have some residual cutoff dependence.
It is however already weak for currently available values of $B_{max}$ and for a large number
of lower and intermediate states. Nevertheless monitoring this dependence is an important
part of the whole analysis.

\begin{figure}
\centerline{\psfig{file=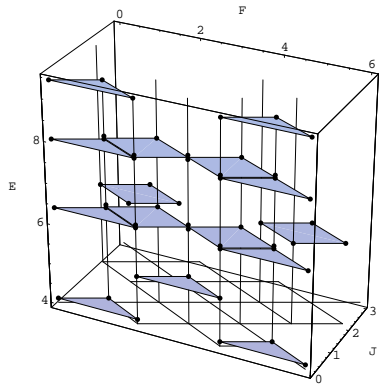,width=5cm}} \vspace*{8pt}
\caption{A sample of dynamical supermultiplets identified for different gluino numbers $F$
 and angular momenta $J$.}
\label{splets}
\end{figure}

Combining together the spectrum and susy fractions allowed us to identify about a dozen of
supermultiplets in a range of angular momenta. A sample of results is shown in Fig.~\ref{splets}.
Very satisfactory restoration of supersymmetry in the lower part of the spectrum was found. It is
also confirmed by our results on the Witten index which tends with the increasing cutoff,
to the time independent constant of $1/4$ ~\cite{ST,SM}. Interestingly the
splitting {\em between} some of the supermultiplets is very small
and has been resolved only with the recent very precise data ~\cite{CB}. Apparently the system is rich enough
that the energy differences between adjacent levels can vary by 2-3 orders of magnitude.

\section{Summary}
Supersymmetric Yang-Mills quantum mechanics is a good laboratory for modern model building. In
the simplest, two space-time dimensional case, we have augmented well known, exact solution
with the explicit construction of the supersymmetric vacuum state. Also recently, an
exact restricted Witten index has been derived together with its manifestly supersymmetric
spectral representation.

     In four space-time dimensions a complete spectrum in various channels of fermionic number and
angular momentum is now available. Expected, since a long time, coexistence of the localized and
non-localized states has been explicitly confirmed.  All eigenstates fall into well identifiable
supermultiplets providing the dynamical realization of supersymmetry. There are two supersymmetric
vacua in this model.
They belong to the continuous spectrum of the scattering states extending into flat valleys of the
potential. Not surprisingly their angular momentum $J=0$. They are related by the particle-hole
symmetry and have non-vanishing fermion numbers $F=2$ and $4$ respectively.
An interesting pattern has been found among the scattering states. They exist only in the "central",
c.f. Fig.~\ref{MAP}, supermultiplets with $F=2,3,4$ and only with {\em even} angular momentum of
bosonic states. Some supermultiplets are almost, but not exactly, degenerate
prompting new interesting questions.

      Due to the lack of space we only signal recent developments in the $D=10$ model~\cite{JW3}. In addition to
 the obvious combinatorial complexity (the system has 51 degrees of freedom),
simultaneous Majorana-Weyl conditions for supersymmetric gluinos result in an unusual situation where
the fermionic number is not conserved, not only by the Hamiltonian, but also by Spin(9) rotations.
In particular, in contradistinction to lower dimensions, the usual empty state is not rotationally invariant.
It turns out that the simplest (no bosons) singlet state is in the most complex sector with twelve fermions.
This state has recently been explicitly constructed and can now be used as the root for the subsequent
building of the whole basis.

\section*{Acknowledgments}

I thank Gabriele Veneziano for interesting discussions. Most of the results presented here were obtained
in the collaboration with Massimo Campostrini.  I also thank the
Theory Group of the Max-Planck-Institute, Munich, for their hospitality. This work is supported by
the Polish Committee for Scientific Research under the grant no PB
1 P03B 024 27 (2004-2007).

\end{document}